\documentclass{article}

\PassOptionsToPackage{compress}{natbib}

\usepackage[preprint]{neurips_2024}

\usepackage[utf8]{inputenc} 
\usepackage[T1]{fontenc}    
\usepackage{hyperref}       
\usepackage{url}            
\usepackage{booktabs}       
\usepackage{amsfonts}       
\usepackage{nicefrac}       
\usepackage{microtype}      
\usepackage{xcolor}         
\usepackage{graphicx}
\usepackage{enumitem}

\title{A Metric for MLLM Alignment in Large-scale Recommendation}

\author{%
  Yubin Zhang*, Yanhua Huang*, Haiming Xu, Mingliang Qi \\
  \texttt{\{zhangyubin, yanhuahuang, xuhaiming, mqi\}@xiaohongshu.com} \\
  \AND
  Chang Wang, Jiarui Jin, Xiangyuan Ren, Xiaodan Wang, Ruiwen Xu \\
  \texttt{\{wangchang2, jinjiarui, renxiangyuan, xiaodan2, ruiwenxu\}@xiaohongshu.com} \\
  \\
  **Equal contribution\\
  \\
  Xiaohongshu Inc.\\
  Shanghai, China \\
}

\begin{document}

\maketitle

\begin{abstract}
  Multimodal recommendation has emerged as a critical technique in modern recommender systems, leveraging content representations from advanced multimodal large language models (MLLMs). To ensure these representations are well-adapted, alignment with the recommender system is essential. However, evaluating the alignment of MLLMs for recommendation presents significant challenges due to three key issues: (1) static benchmarks are inaccurate because of the dynamism in real-world applications, (2) evaluations with online system, while accurate, are prohibitively expensive at scale, and (3) conventional metrics fail to provide actionable insights when learned representations underperform. To address these challenges, we propose the Leakage Impact Score (LIS), a novel metric for multimodal recommendation. Rather than directly assessing MLLMs, LIS efficiently measures the upper bound of preference data. We also share practical insights on deploying MLLMs with LIS in real-world scenarios. Online A/B tests on both Content Feed and Display Ads of Xiaohongshu's Explore Feed production demonstrate the effectiveness of our proposed method, showing significant improvements in user spent time and advertiser value.
\end{abstract}

\section{Introduction}
With increasing user engagement in exploring content feeds, multimodal recommendation techniques play a vital role in modern recommender systems, as they can leverage rich information from content beyond user behaviors. 
Prior works have shown that combining multimodal content representations with user behavior data leads to substantial gains in several applications~\citep{huang2021sliding,zhou2023bootstrap,liu2024alignrec,deng2024end,zhang2025notellm}.

The rapid advancement of multimodal recommendation has been paralleled by remarkable progress in multimodal large language models (MLLMs), such as GPT-4V~\citep{achiam2023gpt}, Gemini~\citep{team2023gemini}, and Qwen-VL~\citep{bai2025qwen2}. 
One of the key lessons in creating state-of-the-art MLLMs is their alignment with human preferences. 
Typically, the alignment is assessed using sophisticated benchmarks~\citep{xuan2025mmlu,zhou2024mlvu,wang2024measuring}, where a weighted average of scores serves as the evaluation metric~\citep{team2023gemini,bai2025qwen2}. While this metric works well for world knowledge domains, its applicability to recommender systems is inherently limited by the system's dynamism: shifting user interests and continuous algorithmic updates. For example, if the current recommender system already incorporates an MLLM's representations, similar representations will fail to deliver significant performance improvements.

To measure the alignment of an MLLM for the current recommender system, the AUC Improvement Score (AIS), defined as the AUC gain when applying MLLM's representation to the ranking model, is commonly adopted as the metric in practice~\citep{sheng2024enhancing,huang2025towards}. While AIS addresses system dynamism by leveraging recent behavior data and the production ranking model, it suffers from the substantial computation costs at scale, primarily due to its dependency on training MLLMs aligned with ranking models and inferring representations on billions of multimodal items. Furthermore, when AIS indicates marginal improvement and further optimization is required, it introduces a diagnostic challenge to distinguish whether the bottleneck lies in (1) suboptimal representation alignment or (2) ineffective utilization of existing representations.

To overcome these challenges, we propose the \textbf{Leakage Impact Score} (LIS), a novel metric that evaluates the quality of preference data construction rather than directly assessing MLLMs. LIS quantifies the ranking performance gap between models trained with and without leaked preference information. Unlike AIS, LIS eliminates the need for expensive MLLM training and inference on multimodal data. We argue that \textbf{LIS can accurately measure the upper bound of preference data, while the capabilities of MLLMs determine how closely we can approach this bound}. Moreover, we share our practical experience on validating preference data using LIS, resulting in effective preference data construction methods.

We conducted online A/B tests in the Explore Feed of Xiaohongshu (also known as the RedNote)~\footnote{https://www.xiaohongshu.com/explore}, to further study LIS. Specifically, we first align MLLMs with preference data validated by LIS and then feed the aligned representation into the production ranking model. The results of the online A/B test in both recommendation and advertisement scenarios demonstrate significant improvements in core metrics, such as user time spent and advertiser value.

Our contributions are summarized as follows: 
\begin{itemize}[leftmargin=0.7cm]
    \item We introduce the Leakage Impact Score (LIS), a novel metric that assesses the upper bound of given preference data, for multimodal recommendation.
    \item Practical experience is shared, where we introduce two types of preference data and demonstrate how to validate them with LIS efficiently.
    \item We conduct online A/B tests on two real-world applications, showing the effectiveness of our proposed methods in large-scale scenarios.
\end{itemize}

\section{Preliminary and Motivation}
This section outlines the standard pipeline for deploying MLLMs aligned with industrial recommender systems, highlighting the complexities and challenges of multimodal recommendation at scale.

As mentioned above, static benchmarks commonly used for evaluating world knowledge domains are not suitable for recommender systems due to their dynamism nature. Consequently, the AUC improvement score (AIS) on the most recent ranking model is adopted in practice, as it demonstrates consistent correlation between offline improvements and online performance. The standard deployment pipeline consists of three key steps:
\begin{itemize}
    \item[Step 1.] \textit{Preference Data Construction}: Prior works have established the critical role of alignment and proposed various data construction methods that leverage user behaviors for preference alignment~\citep{huang2021sliding,liu2024alignrec,zhou2023bootstrap}. This step involves significant effort in preference data design, including cleaning and curation.
    \item[Step 2.] \textit{MLLM Training}: Using the constructed preference data, the next step involves fine-tuning MLLMs to achieve strong validation performance. Prior works have demonstrated the importance of preserving MLLMs' world knowledge while adapting them to recommendation tasks~\citep{yu2025you,wang2025user,zhang2025notellm}. This step requires substantial computational resources for MLLM refinement.
    \item[Step 3.] \textit{Production Model Validation}. The final step evaluates the aligned representations by integrating them into the production model. Prior studies indicate that effectiveness depends heavily on how downstream models utilize these representations~\citep{xing2025esans,sheng2024enhancing,deng2024end}. Therefore, this step involves both algorithmic exploration of representation application and computational overhead for inferring representations.

\end{itemize}
Note that the above three steps form a single iteration. When AIS indicates insufficient improvement, we need to diagnose the issue and repeat the iteration. In practice, this iterative process often requires multiple rounds of execution, creating a bottleneck for further applications of multimodal recommendation.

In the aforementioned pipeline, except for the essential work of human design, we observe substantial computational overhead, particularly when the final AIS performance is unsatisfactory. We identify the root cause as the inability of current methods to properly evaluate the relationship between preference data and the ranking model requirements. A metric to pre-validate preference data could allow the pipeline to focus exclusively on MLLM refinement and downstream application, significantly mitigating the overhead from multiple iterations.

\section{Leakage Impact Score}
\label{sec:lis}
We introduce the \textbf{Leakage Impact Score} (LIS), a novel metric that leverages the concept of data leakage to measure preference data. Data leakage occurs when information from outside the training is involved in the training procedure. Our work focuses specifically on temporal leakage—for instance, when predicting yesterday's behavior while inadvertently including today's data in training.
This example mirrors the real-world constraint where production systems cannot access a user's future interests, making the trained model overestimated~\citep{xin2023user,ji2023critical}.

While data leakage is typically avoided, we repurpose this phenomenon constructively. In recommender systems, models trained with leaked data exhibit offline performance that fails to generalize to online deployment. This discrepancy arises because the online system cannot access the leaked data before inference. 
Note that data unavailable in online deployment may still serve a constructive purpose in offline settings: it provides a mechanism to quantify data importance—irrelevant leaks cause negligible impact, while informative ones lead to significant overestimation.

To this end, we define the LIS as the impact when involving temporally leaked information in the model. The model here refers to the recommender, not the MLLM, thus avoiding the computational overhead associated with MLLMs. In particular, we construct features from leaked data, and the AUC improvement of applying these features to the production ranking model is adopted as the LIS. Note that LIS introduces the upper bound of the preference data, as it is equivalent to a means of accurately predicting future behaviors. If we could validate the effectiveness of an approach to construct the preference data, we believe that MLLMs are able to learn generalized patterns from it.

Here is an example to demonstrate how to calculate LIS in practice. Consider click-through rate prediction, where the production model predicts a user's click probability given their history and a candidate item.
If we augment this model by incorporating the next item clicked by the user as an additional feature to the production model, the resulting AUC improvement constitutes the LIS.
A high LIS suggests this click behavior contains valuable signal for preference data construction. However, since user behaviors are inherently noisy, effectively distilling this signal into MLLM training remains challenging—a challenge we address in the following section.

\section{Practical Experiences}
\label{sec:pe}
In this section, we introduce two types of preference data and how to validate them with LIS. The first type of preference data is sparse representations learned by the recommender. In particular, without loss of generality, we consider the embedding of item ID in the ranking model. If there are multiple embedding slots representing items, we can choose the most important one through the feature importance techniques. We validate the impact of leaked item ID embedding as follows. Formally, let $\mathcal{M}_{T}$ denote the production ranking model serving online at date $T$, i.e., the model have never seen behaviors on date $T$ or thereafter. We replace its item ID embedding with those from $\mathcal{M}_{T+n}$, where $T+n$ denotes the $n$-th date after date $T$. As shown in Table~\ref{tab:id_leak}, this substituion yields LIS values of 0.06 and 0.09 on Xiaohongshu's Explore Feed ranking model. In our settings, where an absolute increase of 0.0010 in AUC is considered significant, these results clearly demonstrate the potential effectiveness of ID embeddings as preference data. 

We attribute this phenomenon to the fact that the item ID serves as the unique identifier for an item, causing the recommender to encode the item's distinctive and important information within its ID embedding. The next section will show that MLLMs can learn generalized information from ID embeddings and gain significant improvement in online experiments.

\begin{table}[htbp]
    \centering
        \begin{tabular}{c|cc}
        \toprule
             & n = 7 & n = 30 \\
        \midrule
            LIS & +0.06 & +0.09 \\
        \bottomrule
        \end{tabular}
    \vspace{0.1in}
    \caption{LIS by leaked ID embedding using the data from $n$ days later.}
    \label{tab:id_leak}
\end{table}

The second type of preference data is from the retrieval perspective. For each item, we identify its 5 most similar items as its side information. This side information is considered as the feature of the target item. Note that if the target item is in the cold-start phase, incorporating similar items with well-learned representations as auxiliary inputs may enhance the cold-start performance. With leaked data, we can accurately identify items similar to a cold-start item with the help of behavior data~\citep{yang2020large}, since we already have knowledge of posterior behaviors. 

However, the results show negligible LIS improvement. This suggests the potential existence of analogous information within the current ranking model. This finding is particularly insightful as it reveals which preference data types merit MLLM-based learning versus those that can be effectively handled by existing system components.

\section{Related Work}
\subsection{MLLM Evaluation}
The primary design objective for MLLMs is to create intelligent chatbots that can thoroughly address human queries spanning both perceptual understanding and logical reasoning. To evaluate these comprehensive capabilities, researchers have developed numerous specialized benchmarks across world knowledge domains. Early studies proposed evaluating MLLMs with visual understanding tasks~\citep{goyal2017making, gurari2018vizwiz}, extend by following works on multilingual~\citep{xuan2025mmlu,tang2024mtvqa},  video understanding~\citep{zhou2024mlvu,fu2025video}, and mathematics~\citep{wang2024measuring,qiao2024we}. Beyond general capabilities, researchers have also explored how to evaluate MLLMs for specific downstream tasks~\citep{hu2024omnimedvqa,wang2024embodiedscan,lu2025can,gao2024evaluating,yu2025benchmarking}, focusing more on the mastery of domain knowledge and skills. However, all these benchmarks employ static construction methodologies, making them unable to generalize to recommender system scenarios—which are inherently dynamic systems where user interests constantly evolve and algorithmic upgrades occur continuously.

\subsection{Multimodal Recommendation}
Multimodal recommendation aims to leverage multimodal representations to improve the recommendation performance, which is critical in modern recommender systems, especially for multi-media content scenarios like TikTok and RedNote. Early works~\citep{he2016vbpr,wei2019mmgcn} only considered the utilization of multimodal representations, ignoring the alignment between multimodal models and recommender systems. Recent works have presented various sophisticated strategies for constructing perference data to align MLLMs with recommender systems. CB2CF~\citep{barkan2019cb2cf} is the first work that considers this alignment by incorporating users' behaviors, where the content encoder learns human preference from collaborative filtering vectors. ~\cite{huang2021sliding} further addressed the instability issue in the learning procedure of original CB2CF and successfully applied it to the diversified recommendation task in a large-scale scenario. ~\cite{zhang2025notellm} proposed maintaining MLLMS' world knowledge capabilities with auxiliary tasks about content predictions. \cite{sheng2024enhancing} argued that MLLMs should learn from deep signals in user behaviors such as search and purchase, proposing to mine hard negatives when constructing negative samples. While the above works have made significant progress in multimodal recommendation, the process of deploying MLLMs for recommendation still suffers from challenges in evaluating the alignment. In this paper, we highlight the challenges and complexities within the multimodal recommendation. To address them, we introduce LIS, which measures the upper bound of given preference data, preventing computational overhead from training MLLMs. Moreover, we also introduce a novel approach that aligns MLLMs with the sparse embeddings learned by the recommender. The online A/B tests on two real-world scenarios demonstrated the effectiveness of our proposed method at scale.

\section{Experiments}
\begin{figure}
    \centering
    \includegraphics[width=\linewidth]{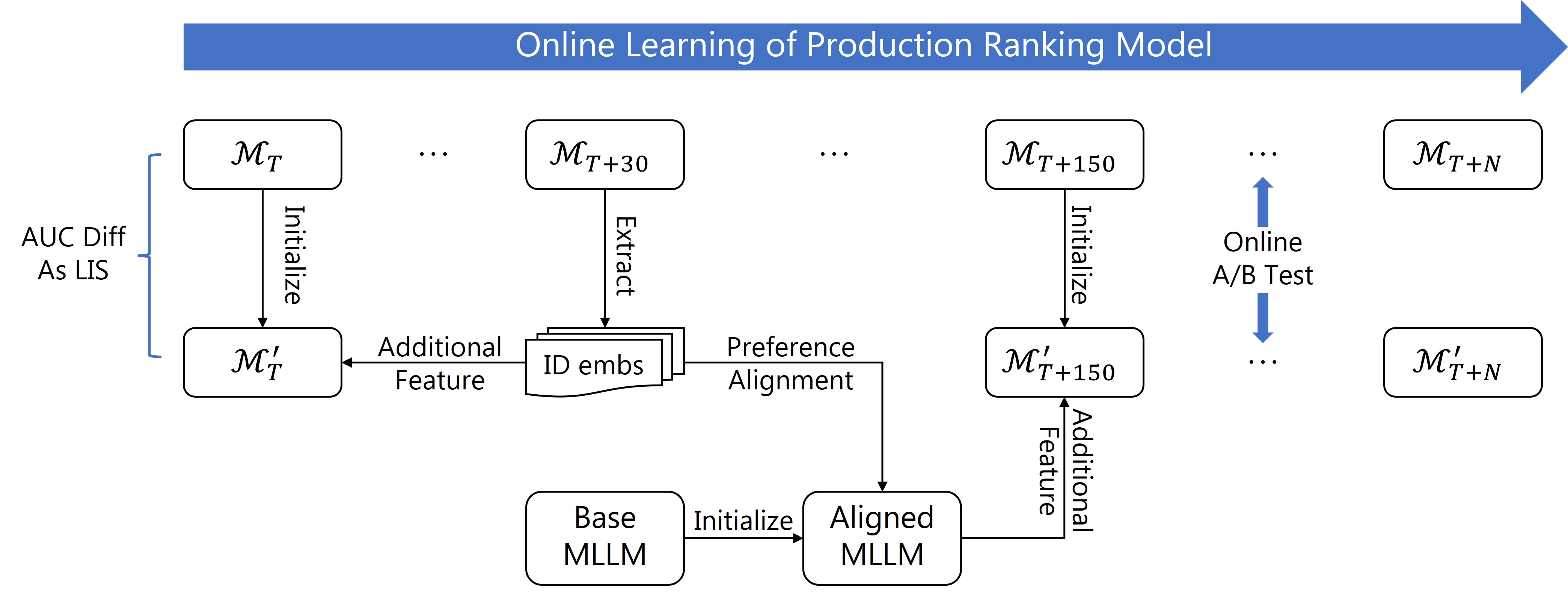}
    \caption{This figure shows how to validate preference data with LIS and how we conduct online A/B tests, where we use ID embeddings as an example of preference data.}
    \label{fig:placeholder}
\end{figure}
\subsection{Implmentation Details}
We conduct large-scale online A/B tests using sparse item ID embeddings as preference data, following the validation described in Section~\ref{sec:pe}. Our experimental setup employs
InternVL~\citep{chen2024internvl} as the base MLLM, with all sparse embeddings extracted from the production ranking model snapshot of May 2024. For data pre-processing, we only retain item embeddings with more than 10000 updates to guarantee statistical reliability. Additionally, we apply data curation inspired by MetaCLIP~\citep{xu2023demystifying}. During MLLM training, we monitor convergence using the mean recall metric on the validation set. 

We evaluate the learned representation in two real-world scenarios, Content Feed and Display Ads, of Xiaohongshu's Explore Feed production. Note that the ranking models for these two scenarios are separate, so we train distinct MLLMs for each scenario. The control group comprises 10\% of randomly selected Xiaohongshu users and applies the production ranking model. For the treatment group, we also randomly select 10\% of users. Each group contains tens of millions of users, with no overlap between groups. Compared to the control group, the ranking model in the treatment group incorporates learned representations as an additional feature, which is the only difference between models in two groups. The added feature—a dense vector with fewer than 100 dimensions—introduces negligible parameter growth, maintaining experimental validity.

\subsection{Online A/B Tests}
For the Content Feed scenario, the experiment was conducted in December 2024. We observe statistically significant improvements across all four key performance metrics: time spent, the number of reads, the number of engagements, and APP lifetime over 30 days (LT30), as presented in Table~\ref{tab:ab_rec}.

\begin{table}[htbp]
    \centering
        \begin{tabular}{c|cccc}
        \toprule
             & Time & Reads & Engagements & LT30 \\
        \midrule
            Improvement & +0.13\% & +0.27\% & +0.40\% & +0.02\% \\
        \bottomrule
        \end{tabular}
    \vspace{0.1in}
    \caption{Online A/B test result in the Content Feed scenario of Xiaohongshu's Explore Feed.}
    \label{tab:ab_rec}
\end{table}

For the Display Ads scenario, the experiment was conducted in November 2024. We observe statistically significant improvements across all four key performance metrics, as shown in Table~\ref{tab:ab_ads}, where Advertiser Value (ADVV) and COST indicate the value of advertisements, while Impression and CTR represent the user experience.

\begin{table}[htbp]
    \centering
        \begin{tabular}{c|cccc}
        \toprule
             & Impression & ADVV & COST & CTR \\
        \midrule
            Improvement & +0.32\% & +0.86\% & +0.79\% & +0.43\% \\
        \bottomrule
        \end{tabular}
    \vspace{0.1in}
    \caption{Online A/B test result in the Display Ads scenario of Xiaohongshu's Explore Feed.}
    \label{tab:ab_ads}
\end{table}

In the aforementioned experiments, we utilized the recommender's representations as preference data and aligned the MLLM with the recommender system through this approach, achieving promising results.
It's worth noting that in our scenario, an item's life-cycle is typically much shorter than 3 months, while the time interval between collecting preference data and conducting online experiments was maintained at least 4 months. That is, none of the items in the preference data appeared as target items during online experiments. Even when directly using the preference data as input for online experiments, no improvement could be obtained.
Therefore, the online improvement primarily stem from the MLLM extracting generalizable patterns from the preference data and successfully applying them to previously unseen items.

\section{Limitations}
While LIS measures the upper bound of given preference data, two challenges remain. First, how to approach the bound, i.e., how to maximize the MLLM's ability to learn the information contained in the preference data. We argue that, in addition to improving the MLLM's general capabilities, potential approaches may include hard mining and curriculum learning. The second challenge concerns how to effectively utilize the learned representations in recommender systems. Since the leaked information carries highly predictive information, ranking models can readily extract useful patterns from it. However, in practical deployment scenarios, the preference data that MLLMs learn from contains no leaked information for online recommenders. Consequently, we contend that the representations obtained through alignment require further investigation of their application methodologies.

\section{Conclusion}
In this paper, we present the leakage impact score (LIS), a novel metric for multimodal recommendation. 
LIS enables assessment of potential effectiveness of preference data before MLLM alignment. Our approach significantly improves deployment efficiency by providing an early-stage validation mechanism.
We further present practical insights on preference data construction, demonstrating that sparse representations learned by ranking models serve as particularly effective preference data for multimodal recommendation. Online A/B tests conducted on two production scenarios, Content Feed and Display Ads in Xiaohongshu's Explore Feed, demonstrate significant improvements, confirming the practical value of our proposed techniques.

\bibliographystyle{rusnat}
\bibliography{references}

\begin{thebibliography}{34}
\providecommand{\natexlab}[1]{#1}
\providecommand{\EM}{\em}
\providecommand{\RNtxt}{\relax}
\RNtxt{}

\bibitem[Achiam et~al.(2023)J.~Achiam, S.~Adler, S.~Agarwal, L.~Ahmad, I.~Akkaya, F.~L. Aleman, D.~Almeida, J.~Altenschmidt, S.~Altman, S.~Anadkat, et~al.]{achiam2023gpt}
{\EM Achiam Josh, Adler Steven, Agarwal Sandhini, Ahmad Lama, Akkaya Ilge, Aleman Florencia~Leoni, Almeida Diogo, Altenschmidt Janko, Altman Sam, Anadkat Shyamal, others }.
\newblock Gpt-4 technical report \allowbreak\newblock// arXiv preprint arXiv:2303.08774. 2023.

\bibitem[Bai et~al.(2025)S.~Bai, K.~Chen, X.~Liu, J.~Wang, W.~Ge, S.~Song, K.~Dang, P.~Wang, S.~Wang, J.~Tang, et~al.]{bai2025qwen2}
{\EM Bai Shuai, Chen Keqin, Liu Xuejing, Wang Jialin, Ge~Wenbin, Song Sibo, Dang Kai, Wang Peng, Wang Shijie, Tang Jun, others }.
\newblock Qwen2. 5-vl technical report \allowbreak\newblock// arXiv preprint arXiv:2502.13923. 2025.

\bibitem[Barkan et~al.(2019)O.~Barkan, N.~Koenigstein, E.~Yogev, O.~Katz]{barkan2019cb2cf}
{\EM Barkan Oren, Koenigstein Noam, Yogev Eylon, Katz Ori}.
\newblock CB2CF: a neural multiview content-to-collaborative filtering model for completely cold item recommendations \allowbreak\newblock// Proceedings of the 13th ACM Conference on Recommender Systems. 2019.  228--236.

\bibitem[Chen et~al.(2024)Z.~Chen, J.~Wu, W.~Wang, W.~Su, G.~Chen, S.~Xing, M.~Zhong, Q.~Zhang, X.~Zhu, L.~Lu, et~al.]{chen2024internvl}
{\EM Chen Zhe, Wu~Jiannan, Wang Wenhai, Su~Weijie, Chen Guo, Xing Sen, Zhong Muyan, Zhang Qinglong, Zhu Xizhou, Lu~Lewei, others }.
\newblock Internvl: Scaling up vision foundation models and aligning for generic visual-linguistic tasks \allowbreak\newblock// Proceedings of the IEEE/CVF conference on computer vision and pattern recognition. 2024.  24185--24198.

\bibitem[Deng et~al.(2024)X.~Deng, L.~Xu, X.~Li, J.~Yu, E.~Xue, Z.~Wang, D.~Zhang, Z.~Liu, G.~Zhou, Y.~Song, et~al.]{deng2024end}
{\EM Deng Xiuqi, Xu~Lu, Li~Xiyao, Yu~Jinkai, Xue Erpeng, Wang Zhongyuan, Zhang Di, Liu Zhaojie, Zhou Guorui, Song Yang, others }.
\newblock End-to-end training of Multimodal Model and ranking Model \allowbreak\newblock// arXiv preprint arXiv:2404.06078. 2024.

\bibitem[Fu et~al.(2025)C.~Fu, Y.~Dai, Y.~Luo, L.~Li, S.~Ren, R.~Zhang, Z.~Wang, C.~Zhou, Y.~Shen, M.~Zhang, et~al.]{fu2025video}
{\EM Fu~Chaoyou, Dai Yuhan, Luo Yongdong, Li~Lei, Ren Shuhuai, Zhang Renrui, Wang Zihan, Zhou Chenyu, Shen Yunhang, Zhang Mengdan, others }.
\newblock Video-mme: The first-ever comprehensive evaluation benchmark of multi-modal llms in video analysis \allowbreak\newblock// Proceedings of the Computer Vision and Pattern Recognition Conference. 2025.  24108--24118.

\bibitem[Gao et~al.(2024)F.~Gao, H.~Jiang, R.~Yang, Q.~Zeng, J.~Lu, M.~Blum, T.~She, Y.~Jiang, I.~Li]{gao2024evaluating}
{\EM Gao Fan, Jiang Hang, Yang Rui, Zeng Qingcheng, Lu~Jinghui, Blum Moritz, She Tianwei, Jiang Yuang, Li~Irene}.
\newblock Evaluating large language models on wikipedia-style survey generation \allowbreak\newblock// Findings of the Association for Computational Linguistics ACL 2024. 2024.  5405--5418.

\bibitem[Goyal et~al.(2017)Y.~Goyal, T.~Khot, D.~Summers-Stay, D.~Batra, D.~Parikh]{goyal2017making}
{\EM Goyal Yash, Khot Tejas, Summers-Stay Douglas, Batra Dhruv, Parikh Devi}.
\newblock Making the v in vqa matter: Elevating the role of image understanding in visual question answering \allowbreak\newblock// Proceedings of the IEEE conference on computer vision and pattern recognition. 2017.  6904--6913.

\bibitem[Gurari et~al.(2018)D.~Gurari, Q.~Li, A.~J. Stangl, A.~Guo, C.~Lin, K.~Grauman, J.~Luo, J.~P. Bigham]{gurari2018vizwiz}
{\EM Gurari Danna, Li~Qing, Stangl Abigale~J, Guo Anhong, Lin Chi, Grauman Kristen, Luo Jiebo, Bigham Jeffrey~P}.
\newblock Vizwiz grand challenge: Answering visual questions from blind people \allowbreak\newblock// Proceedings of the IEEE conference on computer vision and pattern recognition. 2018.  3608--3617.

\bibitem[He, McAuley(2016)R.~He, J.~McAuley]{he2016vbpr}
{\EM He~Ruining, McAuley Julian}.
\newblock VBPR: visual bayesian personalized ranking from implicit feedback \allowbreak\newblock// Proceedings of the AAAI conference on artificial intelligence.  30, 1. 2016.

\bibitem[Hu et~al.(2024)Y.~Hu, T.~Li, Q.~Lu, W.~Shao, J.~He, Y.~Qiao, P.~Luo]{hu2024omnimedvqa}
{\EM Hu~Yutao, Li~Tianbin, Lu~Quanfeng, Shao Wenqi, He~Junjun, Qiao Yu, Luo Ping}.
\newblock Omnimedvqa: A new large-scale comprehensive evaluation benchmark for medical lvlm \allowbreak\newblock// Proceedings of the IEEE/CVF Conference on Computer Vision and Pattern Recognition. 2024.  22170--22183.

\bibitem[Huang et~al.(2025)Y.~Huang, Y.~Chen, X.~Cao, R.~Yang, M.~Qi, Y.~Zhu, Q.~Han, Y.~Liu, Z.~Liu, X.~Yao, et~al.]{huang2025towards}
{\EM Huang Yanhua, Chen Yuqi, Cao Xiong, Yang Rui, Qi~Mingliang, Zhu Yinghao, Han Qingchang, Liu Yaowei, Liu Zhaoyu, Yao Xuefeng, others }.
\newblock Towards Large-scale Generative Ranking \allowbreak\newblock// arXiv preprint arXiv:2505.04180. 2025.

\bibitem[Huang et~al.(2021)Y.~Huang, W.~Wang, L.~Zhang, R.~Xu]{huang2021sliding}
{\EM Huang Yanhua, Wang Weikun, Zhang Lei, Xu~Ruiwen}.
\newblock Sliding spectrum decomposition for diversified recommendation \allowbreak\newblock// Proceedings of the 27th ACM SIGKDD conference on knowledge discovery \& data mining. 2021.  3041--3049.

\bibitem[Ji et~al.(2023)Y.~Ji, A.~Sun, J.~Zhang, C.~Li]{ji2023critical}
{\EM Ji~Yitong, Sun Aixin, Zhang Jie, Li~Chenliang}.
\newblock A critical study on data leakage in recommender system offline evaluation \allowbreak\newblock// ACM Transactions on Information Systems. 2023. 41, 3. 1--27.

\bibitem[Liu et~al.(2024)Y.~Liu, K.~Zhang, X.~Ren, Y.~Huang, J.~Jin, Y.~Qin, R.~Su, R.~Xu, Y.~Yu, W.~Zhang]{liu2024alignrec}
{\EM Liu Yifan, Zhang Kangning, Ren Xiangyuan, Huang Yanhua, Jin Jiarui, Qin Yingjie, Su~Ruilong, Xu~Ruiwen, Yu~Yong, Zhang Weinan}.
\newblock Alignrec: Aligning and training in multimodal recommendations \allowbreak\newblock// Proceedings of the 33rd ACM International Conference on Information and Knowledge Management. 2024.  1503--1512.

\bibitem[Lu et~al.(2025)Y.~Lu, Y.~Yao, J.~Tu, J.~Shao, Y.~Ma, X.~Zhu]{lu2025can}
{\EM Lu~Yuhang, Yao Yichen, Tu~Jiadong, Shao Jiangnan, Ma~Yuexin, Zhu Xinge}.
\newblock Can lvlms obtain a driver’s license? a benchmark towards reliable agi for autonomous driving \allowbreak\newblock// Proceedings of the AAAI Conference on Artificial Intelligence.  39, 6. 2025.  5838--5846.

\bibitem[Qiao et~al.(2024)R.~Qiao, Q.~Tan, G.~Dong, M.~Wu, C.~Sun, X.~Song, Z.~GongQue, S.~Lei, Z.~Wei, M.~Zhang, et~al.]{qiao2024we}
{\EM Qiao Runqi, Tan Qiuna, Dong Guanting, Wu~Minhui, Sun Chong, Song Xiaoshuai, GongQue Zhuoma, Lei Shanglin, Wei Zhe, Zhang Miaoxuan, others }.
\newblock We-math: Does your large multimodal model achieve human-like mathematical reasoning? \allowbreak\newblock// arXiv preprint arXiv:2407.01284. 2024.

\bibitem[Sheng et~al.(2024)X.-R. Sheng, F.~Yang, L.~Gong, B.~Wang, Z.~Chan, Y.~Zhang, Y.~Cheng, Y.-N. Zhu, T.~Ge, H.~Zhu, et~al.]{sheng2024enhancing}
{\EM Sheng Xiang-Rong, Yang Feifan, Gong Litong, Wang Biao, Chan Zhangming, Zhang Yujing, Cheng Yueyao, Zhu Yong-Nan, Ge~Tiezheng, Zhu Han, others }.
\newblock Enhancing Taobao Display Advertising with Multimodal Representations: Challenges, Approaches and Insights \allowbreak\newblock// Proceedings of the 33rd ACM International Conference on Information and Knowledge Management. 2024.  4858--4865.

\bibitem[Tang et~al.(2024)J.~Tang, Q.~Liu, Y.~Ye, J.~Lu, S.~Wei, C.~Lin, W.~Li, M.~F. F.~B. Mahmood, H.~Feng, Z.~Zhao, et~al.]{tang2024mtvqa}
{\EM Tang Jingqun, Liu Qi, Ye~Yongjie, Lu~Jinghui, Wei Shu, Lin Chunhui, Li~Wanqing, Mahmood Mohamad Fitri Faiz~Bin, Feng Hao, Zhao Zhen, others }.
\newblock Mtvqa: Benchmarking multilingual text-centric visual question answering \allowbreak\newblock// arXiv preprint arXiv:2405.11985. 2024.

\bibitem[Team et~al.(2023)G.~Team, R.~Anil, S.~Borgeaud, J.-B. Alayrac, J.~Yu, R.~Soricut, J.~Schalkwyk, A.~M. Dai, A.~Hauth, K.~Millican, et~al.]{team2023gemini}
{\EM Team Gemini, Anil Rohan, Borgeaud Sebastian, Alayrac Jean-Baptiste, Yu~Jiahui, Soricut Radu, Schalkwyk Johan, Dai Andrew~M, Hauth Anja, Millican Katie, others }.
\newblock Gemini: a family of highly capable multimodal models \allowbreak\newblock// arXiv preprint arXiv:2312.11805. 2023.

\bibitem[Wang et~al.(2025)J.~Wang, Y.~Liu, Y.~Sun, X.~Ma, Y.~Wang, H.~Ma, Z.~Su, M.~Chen, M.~Gao, O.~Dalal, et~al.]{wang2025user}
{\EM Wang Jianling, Liu Yifan, Sun Yinghao, Ma~Xuejian, Wang Yueqi, Ma~He, Su~Zhengyang, Chen Minmin, Gao Mingyan, Dalal Onkar, others }.
\newblock User Feedback Alignment for LLM-powered Exploration in Large-scale Recommendation Systems \allowbreak\newblock// arXiv preprint arXiv:2504.05522. 2025.

\bibitem[Wang et~al.(2024{\natexlab{a}})K.~Wang, J.~Pan, W.~Shi, Z.~Lu, H.~Ren, A.~Zhou, M.~Zhan, H.~Li]{wang2024measuring}
{\EM Wang Ke, Pan Junting, Shi Weikang, Lu~Zimu, Ren Houxing, Zhou Aojun, Zhan Mingjie, Li~Hongsheng}.
\newblock Measuring multimodal mathematical reasoning with math-vision dataset \allowbreak\newblock// Advances in Neural Information Processing Systems. 2024{\natexlab{a}}. 37. 95095--95169.

\bibitem[Wang et~al.(2024{\natexlab{b}})T.~Wang, X.~Mao, C.~Zhu, R.~Xu, R.~Lyu, P.~Li, X.~Chen, W.~Zhang, K.~Chen, T.~Xue, et~al.]{wang2024embodiedscan}
{\EM Wang Tai, Mao Xiaohan, Zhu Chenming, Xu~Runsen, Lyu Ruiyuan, Li~Peisen, Chen Xiao, Zhang Wenwei, Chen Kai, Xue Tianfan, others }.
\newblock Embodiedscan: A holistic multi-modal 3d perception suite towards embodied ai \allowbreak\newblock// Proceedings of the IEEE/CVF Conference on Computer Vision and Pattern Recognition. 2024{\natexlab{b}}.  19757--19767.

\bibitem[Wei et~al.(2019)Y.~Wei, X.~Wang, L.~Nie, X.~He, R.~Hong, T.-S. Chua]{wei2019mmgcn}
{\EM Wei Yinwei, Wang Xiang, Nie Liqiang, He~Xiangnan, Hong Richang, Chua Tat-Seng}.
\newblock MMGCN: Multi-modal graph convolution network for personalized recommendation of micro-video \allowbreak\newblock// Proceedings of the 27th ACM international conference on multimedia. 2019.  1437--1445.

\bibitem[Xin et~al.(2023)X.~Xin, J.~Yang, H.~Wang, J.~Ma, P.~Ren, H.~Luo, X.~Shi, Z.~Chen, Z.~Ren]{xin2023user}
{\EM Xin Xin, Yang Jiyuan, Wang Hanbing, Ma~Jun, Ren Pengjie, Luo Hengliang, Shi Xinlei, Chen Zhumin, Ren Zhaochun}.
\newblock On the user behavior leakage from recommender system exposure \allowbreak\newblock// ACM Transactions on Information Systems. 2023. 41, 3. 1--25.

\bibitem[Xing et~al.(2025)H.~Xing, K.~Matsuyama, H.~Deng, J.~Hu, Y.~Zhang, X.~Zeng]{xing2025esans}
{\EM Xing Haibo, Matsuyama Kanefumi, Deng Hao, Hu~Jinxin, Zhang Yu, Zeng Xiaoyi}.
\newblock ESANS: Effective and Semantic-Aware Negative Sampling for Large-Scale Retrieval Systems \allowbreak\newblock// Proceedings of the ACM on Web Conference 2025. 2025.  462--471.

\bibitem[Xu et~al.(2023)H.~Xu, S.~Xie, X.~E. Tan, P.-Y. Huang, R.~Howes, V.~Sharma, S.-W. Li, G.~Ghosh, L.~Zettlemoyer, C.~Feichtenhofer]{xu2023demystifying}
{\EM Xu~Hu, Xie Saining, Tan Xiaoqing~Ellen, Huang Po-Yao, Howes Russell, Sharma Vasu, Li~Shang-Wen, Ghosh Gargi, Zettlemoyer Luke, Feichtenhofer Christoph}.
\newblock Demystifying clip data \allowbreak\newblock// arXiv preprint arXiv:2309.16671. 2023.

\bibitem[Xuan et~al.(2025)W.~Xuan, R.~Yang, H.~Qi, Q.~Zeng, Y.~Xiao, A.~Feng, D.~Liu, Y.~Xing, J.~Wang, F.~Gao, et~al.]{xuan2025mmlu}
{\EM Xuan Weihao, Yang Rui, Qi~Heli, Zeng Qingcheng, Xiao Yunze, Feng Aosong, Liu Dairui, Xing Yun, Wang Junjue, Gao Fan, others }.
\newblock Mmlu-prox: A multilingual benchmark for advanced large language model evaluation \allowbreak\newblock// arXiv preprint arXiv:2503.10497. 2025.

\bibitem[Yang et~al.(2020)X.~Yang, Y.~Zhu, Y.~Zhang, X.~Wang, Q.~Yuan]{yang2020large}
{\EM Yang Xiaoyong, Zhu Yadong, Zhang Yi, Wang Xiaobo, Yuan Quan}.
\newblock Large scale product graph construction for recommendation in e-commerce \allowbreak\newblock// arXiv preprint arXiv:2010.05525. 2020.

\bibitem[Yu et~al.(2025{\natexlab{a}})K.~Yu, R.~Yang, J.~Liao, S.~Li, H.~Li, I.~Li, Y.~Peng, R.~Kamaleswaran, N.~Liu]{yu2025benchmarking}
{\EM Yu~Kunyu, Yang Rui, Liao Jingchi, Li~Siqi, Li~Huitao, Li~Irene, Peng Yifan, Kamaleswaran Rishikesan, Liu Nan}.
\newblock Benchmarking Foundation Models with Multimodal Public Electronic Health Records \allowbreak\newblock// arXiv preprint arXiv:2507.14824. 2025{\natexlab{a}}.

\bibitem[Yu et~al.(2025{\natexlab{b}})Q.~Yu, X.~Wang, S.~Liu, Y.~Bai, X.~Yang, X.~Wang, C.~Meng, S.~Wu, H.~Yang, H.~Xiao, et~al.]{yu2025you}
{\EM Yu~Qing, Wang Xiaobei, Liu Shuchang, Bai Yandong, Yang Xiaoyu, Wang Xueliang, Meng Chang, Wu~Shanshan, Yang Hailan, Xiao Huihui, others }.
\newblock Who You Are Matters: Bridging Topics and Social Roles via LLM-Enhanced Logical Recommendation \allowbreak\newblock// arXiv preprint arXiv:2505.10940. 2025{\natexlab{b}}.

\bibitem[Zhang et~al.(2025)C.~Zhang, H.~Zhang, S.~Wu, D.~Wu, T.~Xu, X.~Zhao, Y.~Gao, Y.~Hu, E.~Chen]{zhang2025notellm}
{\EM Zhang Chao, Zhang Haoxin, Wu~Shiwei, Wu~Di, Xu~Tong, Zhao Xiangyu, Gao Yan, Hu~Yao, Chen Enhong}.
\newblock Notellm-2: Multimodal large representation models for recommendation \allowbreak\newblock// Proceedings of the 31st ACM SIGKDD Conference on Knowledge Discovery and Data Mining V. 1. 2025.  2815--2826.

\bibitem[Zhou et~al.(2024)J.~Zhou, Y.~Shu, B.~Zhao, B.~Wu, S.~Xiao, X.~Yang, Y.~Xiong, B.~Zhang, T.~Huang, Z.~Liu]{zhou2024mlvu}
{\EM Zhou Junjie, Shu Yan, Zhao Bo, Wu~Boya, Xiao Shitao, Yang Xi, Xiong Yongping, Zhang Bo, Huang Tiejun, Liu Zheng}.
\newblock Mlvu: A comprehensive benchmark for multi-task long video understanding \allowbreak\newblock// arXiv e-prints. 2024.  arXiv--2406.

\bibitem[Zhou et~al.(2023)X.~Zhou, H.~Zhou, Y.~Liu, Z.~Zeng, C.~Miao, P.~Wang, Y.~You, F.~Jiang]{zhou2023bootstrap}
{\EM Zhou Xin, Zhou Hongyu, Liu Yong, Zeng Zhiwei, Miao Chunyan, Wang Pengwei, You Yuan, Jiang Feijun}.
\newblock Bootstrap latent representations for multi-modal recommendation \allowbreak\newblock// Proceedings of the ACM web conference 2023. 2023.  845--854.

\end{thebibliography}
\end{document}